\documentstyle[eqsecnum,aps,prd,epsf]{revtex} 
\begin{document}
\def\abstract#1{\begin{center}{\large Abstract}\end{center}
\par #1}
\def\title#1{\begin{center}{{\large #1}}\end{center}}
\def\author#1{\begin{center}{\sc #1}\end{center}}
\def\address#1{\begin{center}{\it #1}\end{center}}
\hfill
\par
\vspace{5mm}
\title{\bf {Strong Cosmic Censorship and Causality Violation}}
\vskip 7mm
\author{ Kengo Maeda \footnote{e-mail:maeda@th.phys.titech.ac.jp; JSPS fellow} 
and Akihiro Ishibashi \footnote{e-mail:akihiro@th.phys.titech.ac.jp}}
\address{Department of Physics, Tokyo Institute of \\ Technology,
Oh-Okayama Meguro-ku, Tokyo 152, Japan}
\vskip 7mm 
\abstract{
We investigate the instability of the Cauchy horizon caused by causality 
violation in the compact vacuum universe with the topology 
$B\times\,{\bf S}^{1}\times {\bf R}$, which Moncrief and Isenberg considered. 
We show that if the occurrence of curvature 
singularities are restricted to the boundary of causality violating region, 
the whole segments of the boundary become curvature singularities.  
This implies that the strong cosmic censorship holds in 
the spatially compact vacuum space-time in the case of the causality 
violation. This also suggests that causality violation cannot occur 
for a compact universe. 
}
\\ 
\\ 
Pacs number(s): 04.20.-q 04.20.Gz 04.20.Dw 98.80.Hw 

\section{Introduction} 

Whether space-time is allowed to have causality violating regions or not is 
an important problem in classical general relativity. 
Tipler~\cite{T1,T2} showed that any attempt to evolve closed 
timelike curves from an initial regular Cauchy data would cause 
singularities to form in a space-time. 
Kriele~\cite{K} showed that the causality violating set has incomplete 
null geodesics if its boundary is compact. Maeda and Ishibashi~\cite{MIa} 
showed that singularities necessarily occur when a boundary of causality 
violating set exists in a space-time under the physically 
suitable assumptions. Thus, over the past few decades a considerable 
number of studies have been made to elucidate the relations between 
causality violation and singularities. 

The appearance of causality violation causes the Cauchy horizon. 
This is closely related to the strong cosmic censorship. 
Its mathematical description is that the maximal globally hyperbolic 
development of a generic initial Cauchy data is inextendible~\cite{ME}. 
In other words, it holds if a space-time admits no Cauchy horizon. 
So far only few attempts have been made to study the instability of 
the Cauchy horizon due to causality violation in classical general relativity. 

One can observe the appearance of causality violations and associated Cauchy 
horizons in compact universe models by extending the space-time maximally. 
The Taub-NUT universe is one of such models. 
In the compact homogeneous vacuum universe case, 
Chru\'sciel and Rendall~\cite{CR} showed that the strong cosmic 
censorship holds in a few class of Bianchi types. 
Ishibashi et al.~\cite{IKSK} showed that the strong cosmic censorship holds 
in compact hyperbolic inflationary universe models. 

On the other hand Moncrief and Isenberg~\cite{MI} showed that 
causality violating cosmological solutions of Einstein equations
are essentially artifacts of symmetries. They proved that there exists 
a Killing symmetry in the direction of the null geodesic generator on 
the Cauchy horizon if the Cauchy horizon is compact by using Einstein
equations. We can easily understand this curious result by 
inspecting exact solutions which have causality violating regions. 
For example, the Misner space-time and the Taub-NUT universe, 
which have compact Cauchy horizons, have Killing symmetries on 
the Cauchy horizons. However, the physically realistic universe is 
inhomogeneous and does not admit Killing symmetries. 
Thus we expect that compact inhomogeneous universe does not have 
any Cauchy horizon or, 
if it does, the Cauchy horizon cannot be compact from the results 
of Moncrief and Isenberg. 
One often studies inhomogeneous universe by adding perturbations 
on homogeneous models. Especially there are some works on the perturbative 
analysis of spatially compact universe with a compact Cauchy horizon. 
As one of such perturbative approaches, Konkowski and Shepley~\cite{KS} 
studied two dimensional cylindrical vacuum space-times. 
They demonstrated the tendency of appearing of scalar curvature 
singularities on the Cauchy horizon (see also Ref.\cite{MTB}). 
These investigations suggest that, if spatially compact universe 
has a Cauchy horizon which divides the space-time into causality preserving 
region and violating region, scalar curvature singularities occur 
somewhere on the Cauchy horizon. 

In this paper we present a theorem as our main result in which if spatially 
compact universe satisfies generic conditions and, if any, 
all the occurring curvature singularities are restricted to the 
boundaries of causality violating regions, then the whole segments 
of the Cauchy horizon become curvature singularities. 
Consequently it follows that such a universe cannot be extended to
the causality violating regions. 

In the next section, we shall classify Cauchy horizons in the compact universe 
into two categories and introduce the definitions for discussing causal 
structure and singularities. In section 3, we briefly review the result of 
Moncrief and Isenberg~\cite{MI}, by introducing Gaussian null coordinates. 
We also review the dual null formalism of Hayward~\cite{H} 
for proving our theorem. We present our theorem in section 4. 
Section 5 is devoted to summary and discussion on the strong curvature 
singularity. 
  
\section{Chronological Cauchy horizon}

In general, one can classify the Cauchy horizons in
the spatially compact space-times into the following two types.
Fig. 1 (a) shows the Cauchy horizon which is caused by a singularity. 
Fig. 1 (b) shows the Cauchy horizon which is caused by causality
violation, where the light corn is tipped allowing the existence
of the closed null geodesic generators. 
The other types of Cauchy horizon, for example, that is caused by 
timelike null-infinity, cannot occur in the case of 
spatially compact space-time under consideration.  
The most distinct feature of the two Cauchy horizons is whether the null 
geodesic generators are closed or not. More precisely, in Fig. 1 (a) 
the null geodesic generators have future endpoints but in Fig. 1 (b) 
they do not. Moncrief and Isenberg~\cite{MI} have considered the case 
of Fig. 1 (b). Their geometrical assumption is the following; 
Let the space-time be $(M,g)$ such as $M = \Sigma\times {\bf R}$ 
where $\Sigma$ is a compact three manifold and the metric $g$ is analytic. 
If there exists a Cauchy horizon $H^{+}(S)$ of 
a partial Cauchy surface $S$, then $H^{+}(S)$ is a compact null 
embedded hypersurface which is diffeomorphic to $\Sigma$ and the topology 
of $H^{+}(S)$ is $B \times\,{\bf S}^{1}$, where $B$ is a compact spacelike 
two-manifold and the ${\bf S}^{1}$ factor is generated 
by the closed null geodesic generator of $H^{+}(S)$. 
More precisely, they considered such a Cauchy horizon $H^{+}(S)$ with 
a local product bundle structure in the sense that; 
\\ 
{\it 
if $H^+(S)$ contains a closed null geodesic generator $\gamma$, 
there exists an open set $U_{\gamma}$ containing $\gamma$ such that \\ 
(i) $U_{\gamma}\cap H^{+}(S)$ is diffeomorphic to $B_{\gamma} \times 
{\bf S}^1$ for some two-manifold $B_{\gamma}$ and some diffeomorphism 
$\phi_{\gamma}: U_{\gamma}\cap H^{+}(S) \rightarrow B_{\gamma} \times 
{\bf S}^1$, and \\ 
(ii) there is a smooth, surjective map $\pi_{\gamma}: 
B_{\gamma} \times {\bf S}^{1} \rightarrow B_{\gamma}$ 
such that, for any $p\in B_{\gamma}$, 
$B_{\gamma} \times {\bf S}^{1} \approx B_{\gamma}\times \pi^{-1}_{\gamma}(p)$ 
and the fiber $\pi^{-1}_{\gamma}(p)$ is diffeomorphic to a closed null 
generator lying in $U_{\gamma}\cap H^{+}(S)$. 
}

The set of the type $U_{\gamma} \cap H^{+}(S)$ is called 
{\it the elementary region of $H^{+}(S)$}. 

They showed that a compact null hypersurface $H^{+}(S)$
has an analytic Killing field $Y$ which is null and tangents to 
a null geodesic generator of $H^{+}(S)$. This interesting 
fact is due to the compactness of $H^{+}(S)$. 
In generic space-times, however, it seems reasonable to suppose 
that the generic condition~\cite{HE} is satisfied. Namely, there is no 
Killing symmetry. Thus, the $H^{+}(S)$ cannot be compact. 
The non-compactness of $H^{+}(S)$ in the generic compact universe 
can be explained by appearance of curvature singularities.

As discussed in Ref.~\cite{MI}, if $M$ has a compact Cauchy horizon 
$H^{+}(S)$, then $M$ has a b-incomplete curve corresponding
to a singular point which had been left out of space-time. 
This singularity is a quasi-regular singularity.  
In this paper we are not concerned with such singularities but 
curvature singularities. We consider the case that the non-compactness of 
the Cauchy horizon is attributed to curvature singularities. 

In general it is useful for describing singularities and 
causal structures to adopt {\it the b-boundary}. 
Schmidt~\cite{S} has constructed the boundary $\partial$ of $M$ 
which is corresponding to singularities of $(M,g)$ 
by using the b-completeness. His construction is characterized 
by distinguishing between infinity and singular points at a finite 
distance. Hereafter we consider a large space $M^{+} := M \cup \partial$. 

Let us introduce the following definition to treat the non-compact 
Cauchy horizon which is caused by appearance of curvature singularities 
in a spatially compact space-time. 
Here, we should comment on the spatially compact space-time manifold 
$M = \Sigma \times {\bf R}$ considered in this paper. 
We want to consider the case that the compact spacelike three-manifold 
$\Sigma$ has a local product bundle structure defined by Moncrief and 
Isenberg as mentioned above, replacing the closed null geodesic generator 
$\gamma$ in the definition by a closed spacelike curve $L$ 
which generates ${\bf S}^{1}$ factor of $\Sigma$. 
In this sense, we write $\Sigma \approx B \times {\bf S}^{1}$ 
throughout the paper. The simplest case is that $\Sigma$ has a global 
product bundle structure; $\Sigma$ is diffeomorphic to $B \times {\bf S}^{1}$ 
and the fibers coincide with closed spacelike curves lying in $\Sigma$. 
Examples of the type $\Sigma \approx {\bf T}^3 \approx {\bf T}^{2} 
\times {\bf T}^{1}$ were constructed by Moncrief in Ref.~\cite{M}. 
The Taub-NUT universe is the non-trivial case; 
$\Sigma$ is diffeomorphic to ${\bf S}^{3}$ with Hopf fibering 
${\bf S}^{3} \rightarrow {\bf S}^{2}$ and fibers ${\bf S}^{1}$ 
coincide with closed spacelike curves in $\Sigma$. \\ 
\\ 
{\it Definition: Chronological Cauchy horizon}. \\   
{\it 
Consider a space-time $(M,g)$ with a partial Cauchy surface $S$ 
of which Cauchy development $D^{+}(S)$ has compact spatial sections 
$\Sigma \approx B\,\times\,{\bf S}^{1}$, where $B$ is a compact orientable 
two-manifold. We call a Cauchy horizon $H^{+}(S)$ {\it the chronological 
Cauchy horizon} if it satisfies the following conditions. \\
(a) Let $\{q_n \}$ be a sequence of points in $D^{+}(S)$ which converges to 
a point $p$ in $H^{+}(S)$. There exists an infinite sequence $\{ L_n \}$ of 
closed spacelike curves which generate ${\bf S}^1$ factors of $D^{+}(S)$ and 
each $L_n$ passes through $q_n$ such that, for every point $r_n\in L_n$, 
the tangent vector $K_n $ of $L_n$ at $r_n$ approaches to null, i.e. 
\[ \lim_{q_n \to p} g(K_n,K_n)|_{r_n}=0. \]  
\\ 
(b) If $H^+(S)$ contains a closed null geodesic generator $\gamma$, 
there exists an elementary region of the type $U_{\gamma} \cap H^+(S)$ 
with the local product bundle structure as mentioned above. 
\\ 
(c) There exists a compact spacelike orientable two-surface $B$ 
on $H^{+}(S)$ such that there is no null geodesic generator of $H^{+}(S)$
which connects two different points $p$ and $q\,(\neq p)$ on $B$. \\ 
}
  
It is obvious that the chronological Cauchy horizon 
has no future endpoint in $(M,g)$ from the condition (a). 
This means that the segments of ${\dot{D}}^{+}(S,M^{+})$ 
are null and, if exist, singularities are restricted to 
the boundary of causality violating region in $M^{+}$. 
As mentioned above, when we speak of singularity in this paper, 
it means curvature singularity. Thus, hereafter, $\partial$ denotes 
curvature singularity. 

The chronological Cauchy horizon is not required in general to be
compact and is a generalization of the Cauchy horizon which Moncrief
and Isenberg considered.  Indeed, the chronological Cauchy horizon can
be non-compact due to the existence of curvature singularities
$\partial$. In that case, there are non-closed incomplete null
geodesic generators of $H^+(S)$, which terminate at $\partial$, and
the condition (c) implies that such a non-closed null geodesic
generator does not intersect $B$ more than once.  
If $\partial$ is empty, the chronological Cauchy horizon is compact, i.e. 
it can be covered by a finite number of the elementary regions of the type
$U_{\gamma} \cap H^+(S)$, and diffeomorphic to $\Sigma$.

\section{Preliminaries} 
In this section, we introduce the Gaussian null coordinates and 
review Moncrief and Isenberg's theorem for the discussion of our theorem. 
We concentrate our interest on the compact vacuum space-time 
$M \approx \Sigma \times {\bf R}$ which admits the chronological 
Cauchy horizon defined in the previous section. In addition we introduce 
the dual null coordinates~\cite{H} and define a strong curvature singularity 
condition in order to prove our theorem in the next section.  

\subsection{The Gaussian null coordinates.}
We adopt the Gaussian null coordinates $\{t,x^3,x^a\}\, (a=1,2)$ 
in the neighborhood of the chronological Cauchy horizon 
(in detail, see Ref.~\cite{MI}). 
In this coordinate system, the metric takes the following form;
\begin{equation}
g=2dtdx^3+\phi(dx^3)^2+2\beta_a dx^adx^3 +h_{ab}dx^a dx^b. \label{g:GNC}
\end{equation}
The chronological Cauchy horizon $H^{+}(S)$ corresponds to the
hypersurface $t=0$. The future development $D^{+}(S)$ is the region $t<0$. 
In $D^{+}(S)$, the coordinate basis vector ${\partial}/{\partial x^{3}}$ 
has closed spacelike integral curves which generate ${\bf S}^1$ factor 
and $h_{ab}$ is the induced metric of 
the spacelike two-surface $B$. One can choose the coordinates 
$\{t,x^a,x^3\}$ such that $\phi=\beta_a=0$ at $H^{+}(S)$. 
In this coordinates, $R_{33}$ component of the Ricci tensor is given by 
\begin{eqnarray}   
R_{33} &=&  \frac{1}{\sqrt{h}} \left[ \sqrt{h} \left(\frac{1}{2}\phi_{,3}+
\frac{1}{2}\phi\phi_{,t}+\frac{1}{2}\beta^a \phi_{,a}
-\beta^a \beta_{a,3}-\frac{1}{2}\beta^a \beta_a \phi_{,t} \right) \right]_{,t} 
\nonumber \\
&+& \frac{1}{\sqrt{h}}\left[ \sqrt{h}h^{ac} 
\left( -\frac{\phi_{,a}}{2}+\beta_{a,3}+\frac{\beta_a}{2}\phi_{,t} \right) 
\right]_{,c} 
\nonumber \\ 
&-& \frac{1}{\sqrt{h}} \left[ \sqrt{h}\frac{\phi_{,t}}{2} \right]_{,3} 
- \frac{1}{2}h^{ab}h_{ab,33}+\frac{1}{4}h^{ac}h^{bd}h_{ab,3}h_{cd,3} 
\nonumber \\ 
&-& \Bigg[ \frac{1}{2} ( \phi_{,t}-\beta^a \beta_{a,t})^2 
+\frac{1}{4}h^{ac}h^{bd} (\beta_{a,b}-\beta_{b,a})(\beta_{d,c}-\beta_{c,d}) 
\nonumber \\
&+& \frac{1}{2}h^{dc}\beta_{c,t} \left (2\phi_{,d}+\phi \beta_{d,t}-2\beta^a(
\beta_{a,d}-\beta_{d,a})-2\beta_{d,3}-\beta^a \beta_a \beta_{d,t}
\right) \Bigg],  
\label{Ricc:33}  
\end{eqnarray}
where $\beta^a := h^{ab}\beta_b,\,h := \det(h_{ab})$. The other 
components are explicitly given in Ref.~\cite{MI}. 

On the chronological Cauchy horizon $H^{+}(S)$, 
substituting $\phi=\beta_a=0$ into Eq.~(\ref{Ricc:33}), we obtain  
\begin{equation}
\left( \ln{\sqrt{h}} \right)_{,33}+\frac{1}{2}\phi_{,t} \left( \ln{\sqrt{h}} 
\right)_{,3} + \frac{1}{4}h^{ac}h^{bd}h_{ab,3}h_{cd,3}=0.  \label{eq:Ray} 
\end{equation}
In the case that the Cauchy horizon $H^{+}(S)$ is a compact analytic embedded
null hypersurface, applying the maximum principle~\cite{Max}
for Eq.~(\ref{eq:Ray}), we obtain $h_{,3}=0$ and consequently  
\begin{equation}
h_{ab,3}=0,  \label{eq:h3} 
\end{equation}  
by substituting $h_{,3}=0$ into Eq.~(\ref{eq:Ray}) again. 
We can see that Eq.~(\ref{eq:h3}) is identical to the Killing equation 
\begin{equation}
Y_{\nu;\mu}+Y_{\mu;\nu}=0,  \label{eq:Killing}  
\end{equation}
where $Y^{\mu}({\partial}/{\partial x^{\mu}}) := {\partial}/{\partial x^{3}}$ 
and the semicolon represents the covariant derivative with respect to 
the metric $g$ in Eq.~(\ref{g:GNC}).  
Thus we can observe that the compact Cauchy horizon $H^{+}(S)$ has a Killing 
symmetry along the direction of ${\partial}/{\partial x^{3}}$. 
This is the result which Moncrief and Isenberg obtained. 

Next we introduce the dual null coordinates and define
a strong curvature singularity condition which is slightly different 
from Kr\'olak's one in Ref~\cite{KR} and weaker than it. \\
  
\subsection{The dual null coordinates.}
In the dual null coordinates $\{u,v,x^a\}$, the metric is written as 
\begin{equation}
g=-2e^{-\lambda}du dv+s^a s_a du^2+2s_a dx^a du+h_{ab}dx^a dx^b. 
\end{equation}
One can easily understand that the dual null coordinates are 
transformed into the Gaussian null coordinates by taking
\begin{equation}
du=-e^{\lambda}dx^3,\,dv=dt,\,\phi 
= s^a s_a e^{2\lambda},\,\beta_a = -2s_a e^{\lambda}.
\end{equation}
We introduce some quantities, of which notation is as those in Ref.~\cite{BC}. 
Introducing the null vectors 
\begin{equation}
k^{\mu} = (\partial_{v})^{\mu} =(1,0,0,0),\quad 
n^{\mu}=(\partial_{u})^{\mu}=(0,1,-s^{1},-s^{2}), 
\end{equation}
we define
\begin{equation}
\Sigma_{ab} := {\cal L}_k h_{ab},\quad \tilde{\Sigma}_{ab} 
:= {\cal L}_n h_{ab}, 
\end{equation}
where ${\cal L}_k$ represents the Lie derivative along the vector field 
$k^{\mu}$. The expansions $\theta,\,\tilde{\theta}$, 
the shears $\sigma_{ab},\tilde{\sigma}_{ab}$, and the twist vector 
$\omega_a$ are represented as 
\begin{equation}
\theta=\frac{1}{2}h^{ab}{\Sigma}_{ab},\quad \tilde{\theta}=
\frac{1}{2}h^{ab}\tilde{\Sigma}_{ab}, 
\end{equation}
\begin{equation}
\sigma_{ab}=\Sigma_{ab}-\theta h_{ab},\quad \tilde{\sigma}_{ab}
= \tilde{\Sigma}_{ab}-\tilde{\theta}h_{ab}, 
\end{equation}
\begin{equation}
\omega_a=\frac{1}{2}e^{\lambda}h_{ab}{\cal L}_k s^b. 
\end{equation}
On $H^{+}(S)$, the null vector $n^{\mu}$ corresponds to a tangent vector 
of a null geodesic generator of $H^{+}(S)$. 
 
\subsection{The strong curvature singularity.} 
As discussed in the previous section, we want to consider space-times 
which contain the chronological Cauchy horizon $H^+(S)$ and 
curvature singularity restricted to the boundary of causality violating 
region $\overline{H^+(S, M^+)}$. Thus such a singularity can be specified 
by, especially, the incomplete null geodesic generator of the chronological 
Cauchy horizon $H^+(S)$. \\ 
\\ 
{\it Definition: Strong curvature singularity. \\
A future inextendible null geodesic generator $l$ of $H^{+}(S)$ is said  
to terminate in a strong curvature singularity in the future if
there exists a point $p$ on $l$ such that the expansion $\tilde{\theta}|_p$ 
is negative in the future direction.} \\ 

We will discuss whether or not the strong curvature singularity 
can occur in the vacuum space-time by using the dual null formalism 
of Hayward in the last section. 

\section{Theorem}

In this section, we present our theorem in which 
no spatially compact space-time can have a chronological Cauchy horizon 
under the seemingly physical assumptions. 

\begin{quote}
{\bf Theorem}\\
{\it Let $(M,g)$ be a spatially compact vacuum space-time which admits 
a regular partial Cauchy surface $S$ diffeomorphic to 
$\Sigma \approx B \times {\bf S}^1$. If $(M,g)$ satisfies the following 
conditions,\\
(i) the generic condition, i.e. every inextendible null geodesic contains 
a point at which $K_{[a}R_{b]cd[e}K_{f]}K^c K^d\neq 0$, 
where $K^a$ is the tangent vector to the null geodesic, \\
(ii) the Cauchy horizon, if any, is the chronological Cauchy horizon 
$H^{+}(S)$, \\
(iii) all occurring curvature singularities are the strong curvature 
singularities, \\ 
then, $(M,g)$ is globally hyperbolic.} \\
\end{quote}
{\it Proof.} 

Suppose that space-time $(M,g)$ is not globally
hyperbolic. Then either $H^{+}(S)$ or $H^{-}(S)$ exists. 
Let us consider only $H^{+}(S)$ without loss of generality and 
take the dual null coordinates $\{u,v,x^a\}$ defined in the previous section 
in some neighborhood of $H^{+}(S)$. 
The generator of $H^{+}(S)$ has no past endpoint in $(M,g)$ 
since $S$ is a partial Cauchy surface. In addition every null geodesic 
generator of $H^{+}(S)$ has no future endpoint in $(M,g)$ 
from the definition of the chronological Cauchy horizon. 
There exists a point such that $d \tilde{\theta}/{du}\neq 0$ in each 
null geodesic generator from the condition (i) and the Raychaudhuri 
equation of a null geodesic generator of $H^{+}(S)$. 

Suppose that there exists a closed null geodesic generator $\gamma$ of 
$H^+(S)$. There exists an elementary region $U_{\gamma} \cap H^+(S)$ 
which contains $\gamma$ from the condition (b) of the definition of 
the chronological Cauchy horizon. 
Then, Eq.~(\ref{eq:Killing}) is satisfied in the elementary 
region $U_{\gamma} \cap H^+(S)$ from the theorem of Moncrief and 
Isenberg~\cite{MI}. 
This contradicts with $d\tilde{\theta}/{du}\neq 0$. 
Therefore any null geodesic generator of $H^+(S)$ cannot be closed. 
Since $H^+(S)$ is a chronological Cauchy horizon 
and $\overline{H^{+}(S,M^{+})} \cup \partial$ generates ${\bf S}^{1}$ factor, 
every null geodesic generator $l$ of $\overline{H^{+}(S, M^{+})}$ terminates 
at some points of $\partial$ both in the future and past directions. 
In addition, such null geodesic generators do not intersect $B$ 
more than once from the condition (c). 
Here the curvature singularity $\partial$ is a strong curvature singularity 
from the condition (iii). Thus the expansions of the future directed null 
generators of $\overline{H^{+}(S, M^{+})}$ become negative in the future 
direction somewhere near the future endpoints in $\partial$. 
On the other hand the expansions of past directed null generators also 
become negative in the past direction somewhere near the past endpoints 
in $\partial$.    

Let $\Gamma_u$ be a null hypersurface on which $u = \mbox{const.}$ 
with tangent $k^{\mu}$ in the neighborhood and of which intersection with 
$H^{+}(S)$, i.e. $\Gamma_u \cap H^{+}(S)$, coincides with $B$. 
From the ansatz of the dual null formalism, $\Gamma_u$ is foliated 
by compact spacelike two-surfaces $B_m$ and hence we can take an infinite 
sequence of the two-surfaces $\{B_m\}$ on $\Gamma_u$ which converges to $B$. 
Let us consider the boundaries of the causal past sets ${\dot{J}}^{-}(B_m)$.  
For each number $m$, ${\dot{J}}^{-}(B_m)$ is closed due to the compactness 
of the spatial section of $D^{+}(S)$ and hence there exists a null geodesic 
generator $l_{m}$ of ${\dot{J}}^{-}(B_m)$ whose future and past endpoints, 
denoted by $p_m$ and $q_m$ respectively, are on $\Gamma_u$ and the tangent 
vector at $p_m$ is $n^{\mu}$. 
The limit points $p$ and $q$ of the infinite sequences 
$\{p_m\}$ and $\{q_m\}$, respectively, are on $B$. 
Here $q$ does not necessarily coincide with $p$. 
Then, from the limit curve lemma in Ref.~\cite{BM}, 
for the infinite sequence of the null geodesic generators $\{l_{m}\}$, 
there exist limit null geodesic curves $l_p$ and $l_q$ on $H^{+}(S)$ 
which pass through the points $p$ and $q$, respectively and 
a subsequence $\{l_n\}$ which converges to $l_p$ and $l_q$ uniformly 
with respect to $h$. Here $h$ is a complete Riemannian metric 
on the space-time (in detail see Ref.~\cite{BM}). 

As mentioned above, since every null geodesic generator of $H^{+}(S)$ 
terminates at $\partial$, the limit curves $l_p$ and $l_q$ also terminate 
at $\partial$ in the past and future, respectively, 
without intersecting $B$ more than once. 
In addition, because $\partial$ is a strong curvature singularity, 
$l_p$ and $l_q$ have points $t$ and $s$ such that 
$\tilde{\theta}|_t > 0$ and $\tilde{\theta}|_s < 0$ in the future direction.  
Since $l_n$ converges to $l_p$ and $l_q$ uniformly, for any neighborhoods 
$U_s$ and $U_t$ of the points $s$ and $t$ respectively, there exists 
a natural number $N$ such that all $l_n(n>N)$ intersect both $U_{s}$ 
and $U_{t}$. It also can be taken the infinite sequences of 
points $\{s_n\}$ and $\{t_{n}\}$ such that, 
for each $n$, $s_n \in U_s\cap l_n$, 
$t_n \in U_t\cap l_n$, $s_n < t_n$, and these sequences 
converge to $s\in l_q$ and $t\in l_p$, respectively.  
Then, there exists a number $N'$ such that, for any  
$n>N'$, $\tilde{\theta}|_{t_n} > 0$, $\tilde{\theta}|_{s_n} < 0$, in the 
future direction by continuity. 
This contradicts with the fact that the expansion 
of $l_n$ must decrease monotonically in the future direction. 
Consequently, $H^{+}(S)$ cannot exist.   
\hfill$\Box$ 

\section{Conclusion and discussions} 
We showed that in the compact universe, if the curvature singularity is 
restricted to the boundary of causality violating set,
the whole segments of the boundary become curvature singularities. 
Consequently the vacuum space-time with compact spatial section 
$\Sigma \approx B\times {\bf S}^1$ cannot be extended to the 
causality violating region. 
The result means that the strong cosmic censorship holds 
in such a space-time. 

In our proof of the theorem, we use the strong curvature singularity 
condition, whose notion was first introduced by Tipler~\cite{T3} 
and described in terms of expansions by Kr\'olak~\cite{KR}. 
Our definition of the strong curvature singularity is slightly different 
from that by Kr\'olak. Therefore it is worth to discuss whether or not 
our strong curvature singularity condition is reasonable 
in the vacuum space-time. Here we use the dual null coordinates 
of Hayward~\cite{H}. 

Let $\Gamma_v$ be the null hypersurface $v = \mbox{const.}$.  
On $\Gamma_v$, the Raychaudhuri equation can be written by
\begin{equation} 
\frac{d\tilde{\theta}}{du}=-\frac{1}{2}{\tilde{\theta}}^2
-\frac{d\lambda}{du}\tilde{\theta} -\frac{1}{4}
\tilde{\sigma}_{ab}\tilde{\sigma}^{ab}.  \label{eq:Dtilde}
\end{equation}
Hayward noticed that Eq.~(\ref{eq:Dtilde}) may be simplified 
by making use of coordinate freedom on $\Gamma_v$. Choosing the coordinate $u$ 
on $\Gamma_v$ as  
\begin{equation}
\frac{d\lambda}{du}=-\frac{1}{2}\tilde{\theta},  \label{Gauge:u}  
\end{equation}
Eq.~(\ref{eq:Dtilde}) is written by
\begin{equation}
\frac{d\tilde{\theta}}{du}=-\frac{1}{4}\tilde{\sigma}_{ab} 
\tilde{\sigma}^{ab}, \label{eq:tildeshear} 
\end{equation}
The $\omega_a$, $\theta$, and $\tilde{\theta}$ can be easily integrated 
along $\Gamma_{v}$ and $\Gamma_{u}$ respectively by using the vacuum 
Einstein equations (in detail, see Ref.~\cite{BC}) as follows,  
\begin{eqnarray}  
&&{} 
 \omega_a(u) = \omega_a|_o \exp\left[-\int^u_{u_o}du'\tilde{\theta}(u')
\right] +\exp\left[-\int^u_{u_o}du'\tilde{\theta}(u')\right] \nonumber \\
&&{} 
\qquad 
\times  \int^u_{u_o}du'\left(-\frac{1}{2}\bigtriangleup^b\tilde{\sigma}_{ab}
+\frac{3}{4}\bigtriangleup_a \tilde{\theta}
+\frac{1}{2}\tilde{\theta}\bigtriangleup_a \lambda \right)
\exp\left[ \int^{u'}_{u_o} du''\tilde{\theta}(u'') \right], \\
&&{} 
\theta(u) = \theta|_o \exp\left[-\int^u_{u_o}du'\tilde{\theta}(u') \right]
+\exp\left[-\int^u_{u_o}du'\tilde{\theta}(u') \right] 
\times \int^u_{u_o}du' e^{-\lambda}K\exp\left[\int^{u'}_{u_o}du''
\tilde{\theta}(u'')\right], 
\label{theta} \\ 
&&{} 
\tilde{\theta}(v) = \theta|_o \exp\left[-\int^v_{v_o}dv'\theta(v')\right]
+\exp\left[-\int^v_{v_o}dv'\theta(v')\right] 
\times \int^v_{v_o}dv' e^{-\lambda}L\exp\left[\int^{v'}_{v_o}dv''\theta(v'')
\right]. 
\label{tildetheta}  
\end{eqnarray}
Here $K$ and $L$ are defined respectively as
\begin{eqnarray} 
&&{} 
K:=-\frac{1}{2}{}^{(2)}\!R+\omega_a \omega^a 
-\frac{1}{2}\bigtriangleup^a\bigtriangleup_a \lambda
+\frac{1}{4}\bigtriangleup^a \lambda
\bigtriangleup_a \lambda - \omega^a \bigtriangleup_a \lambda
+\bigtriangleup_a \omega^a, \label{K} \\ 
&&{} 
L:=-\frac{1}{2}{}^{(2)}\!R+\omega_a \omega^a
-\frac{1}{2}\bigtriangleup^a\bigtriangleup_a \lambda
+\frac{1}{4}\bigtriangleup^a \lambda
\bigtriangleup_a \lambda + \omega^a \bigtriangleup_a \lambda
-\bigtriangleup_a \omega^a,  \label{L}
\end{eqnarray}
and $\bigtriangleup_a$, ${}^{(2)}\!R$ are, respectively, 
the covariant derivative, Ricci scalar with respect to $h_{ab}$.  

On the other null hypersurface $\Gamma_u$ on which $u = \mbox{const.}$, 
the Raychaudhuri equation is written by
\begin{equation}
\frac{d\theta}{dv}=-\frac{1}{2}{\theta}^2  
-\frac{d\lambda}{dv}\theta -\frac{1}{4} \sigma_{ab}\sigma^{ab}. 
\label{eq:Rayv} 
\end{equation}
As well as on $\Gamma_v$, choosing the coordinate $v$ on $\Gamma_u$
such that
\begin{equation}
\frac{d\lambda}{dv}=-\frac{1}{2}\theta,  \label{Gauge:v}   
\end{equation}
we can rewrite Eq.~(\ref{eq:Rayv}) as 
\begin{equation}
\frac{d\theta}{dv}= -\frac{1}{4}\sigma_{ab}\sigma^{ab}.  \label{eq:shear}  
\end{equation}
With the help of Eqs.~$(\ref{Gauge:u})$ and $(\ref{Gauge:v})$, 
we can express Eqs.~$(\ref{theta})$ and $(\ref{tildetheta})$ 
on each null hypersurface $\Gamma_v$, $\Gamma_u$ such as 
\begin{eqnarray}
\theta(u) &=& \theta_o e^{2(\lambda-\lambda_{u_o})}
+e^{2\lambda}\int^u_{u_o}du'e^{-3\lambda}K,  \label{theta:u}
\\ 
\tilde{\theta}(v) &=& \tilde{\theta}_o e^{2(\lambda-\lambda_{v_o})}
+e^{2\lambda}\int^v_{v_o}dv'e^{-3\lambda}L. \label{theta:v} 
\end{eqnarray}

In the vacuum space-time, the strong curvature singularities are caused 
by Weyl tensor only. The Weyl tensor produces the shear tensor 
$\sigma_{ab}(\tilde{\sigma}_{ab})$ and the square $\sigma_{ab}\sigma^{ab}
(\tilde{\sigma}_{ab}\tilde{\sigma}^{ab})$, which can be interpreted as the 
gravitational energy. In the Kerr black hole case, Brady and 
Chambers~\cite{BC} showed that only the quantity $\sigma_{ab}\sigma^{ab}$ 
diverges on the Cauchy horizon 
but $\tilde{\sigma}_{ab}\tilde{\sigma}^{ab}$ does not. 
In terms of expansions, this means that only the expansion $\theta$ diverges 
but $\tilde{\theta}$ does not. In generic space-times, however, 
it is expected that $\theta$ and $\tilde{\theta}$ 
behave similarly; both $\theta$ and $\tilde{\theta}$ diverge as they approach 
the curvature singularity. 
Indeed, from Eqs.~$(\ref{theta:u})$ and $(\ref{theta:v})$, 
it turns out that, if $\lambda$ diverges, both $\theta$ and $\tilde{\theta}$ 
diverge. If there exists a curvature singularity such that 
$\theta(u)$ diverges while $\tilde{\theta}(v)$ does not, 
then $K$ must diverge but $\lambda$ and $L$ must not diverge. 
In the case, because $L$ is different from $K$ only the signature of the last 
two terms in Eq.~(\ref{L}): 
$+ \omega^a \bigtriangleup_a \lambda -\bigtriangleup_a \omega^a$, 
the divergence of these two terms must cancel out that of all the other terms 
in $L$. Such a case is unlikely and cannot be considered as generic. 
This suggests that, in generic space-times, if at least either 
$\sigma_{ab}\sigma^{ab}$ or $\tilde{\sigma}_{ab}\tilde{\sigma}^{ab}$ diverges 
on the curvature singularity, both $\theta$ and $\tilde{\theta}$ diverge and 
hence our strong curvature singularity condition is satisfied. 
This means that the expansions of the null geodesic generators 
on $H^{+}(S)$ diverge on the singularity whenever the gravitational energy 
diverges. In addition, this suggests that the expansion of each incomplete 
causal geodesic diverges on the singularity independent of its tangent in 
generic vacuum space-times. 
The rigorous study of the discussion above will be given in future works. 

One might consider that there exists a possibility to cause causality 
violation in the presence of matter. However, in the black hole case, 
the existence of matter does not change the property of the Cauchy horizon 
drastically as Brady and Smith~\cite{BS} have shown by numerical 
investigation. Thus we believe that the strong cosmic censorship 
in compact universe also holds even if matter exists. 

\begin{center} {\bf \large Acknowledgements} \end{center}

We would like to thank Professor A.Hosoya for useful discussions and 
helpful suggestions. We are grateful to T.Koike, M.Narita, 
K.Tamai and S.Ding for useful discussions. We are also grateful to 
T.Okamura and K.Nakamura for kind advice and comments. 
K.M. acknowledges financial supports from the Japan Society for the Promotion 
of Science and the Ministry of Education, Science and Culture.

\begin{figure}[htbp]
 \centerline{\epsfxsize=10.0cm \epsfbox{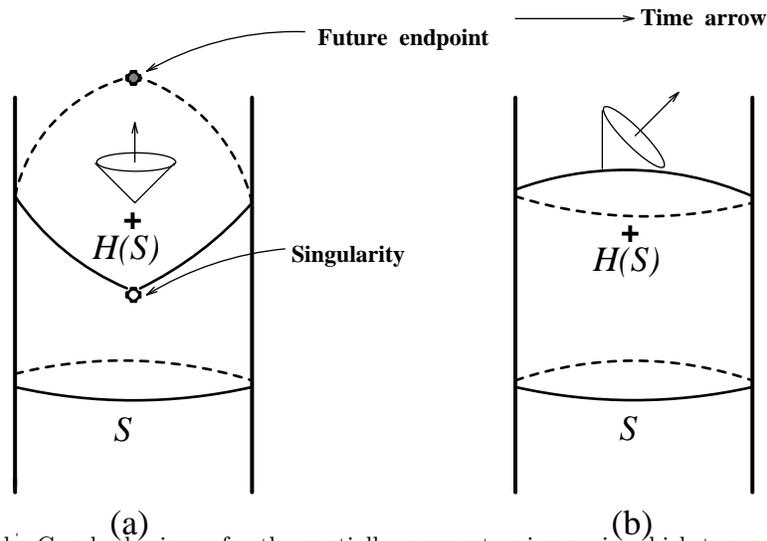}}
        \caption{ Two types of the Cauchy horizons for the spatially 
                  compact universe, in which two-spatial dimensions, 
                  $B$ factors, are suppressed. }
        \protect 
\end{figure}

\end{document}